\documentclass[aps,pre,amsmath,amssymb,twocolumn,showpacs,superscriptaddress,floatfix]{revtex4-1}

\usepackage{graphicx}
\usepackage{dcolumn}
\usepackage{bm}
\usepackage{hyperref}
\usepackage{natbib}
\usepackage{color}

\newcommand{\vect}[1]{{\mathbf{#1}}}
\newcommand{\ind}[1]{{\mathrm{#1}}}
\newcommand{\erf}{\mathrm{erf}}

\begin{document}


\title{Simulation of particle mixing in turbulent channel flow due to intrinsic fluid velocity fluctuation}

\author{Thomas Burgener}
\email[Electronic address: ]{buthomas@ethz.ch}
\author{Dirk Kadau}
\author{Hans J. Herrmann}
\affiliation{Computational Physics, IfB, ETH-H\"onggerberg, Schafmattstrasse 6, 8093 Z\"urich, Switzerland}


\date{\today}

\begin{abstract}
We combine a DEM simulation with a stochastic process to model the movement of spherical particles in a turbulent channel flow. With this model we investigate the mixing properties of two species of particles flowing through the channel. We find a linear increase of the mixing zone with the length of the pipe. Flows at different Reynolds number are studied. Below a critical Reynolds number at the Taylor microscale of around $R_{c}\approx 300$ the mixing rate is strongly dependent on the Reynolds number. Above $R_{c}$ the mixing rate stays nearly constant.
\end{abstract}


\pacs{47.27.T-, 45.70.Mg, 83.50.Xa, 47.27.E-}


\maketitle


\section{\label{sec:intro}Introduction}
Transport of granular material by a fluid plays an important role in many environmental and technological processes. In nature examples are sand storms in the desert, the formation of dunes \cite{sauermann31305,Almeida2006} and river deltas \cite{seybold16804} or the dispersion of particles like pollen or carbon black in the atmosphere or microorganisms or pollution in the ocean \cite{Perlekar2010} or also particle transport and separation in lung-like structures \cite{Vasconcelos2010} and rough channels\cite{Vasconcelos2009}. Industrial and technical applications range from transport of contaminant and pollution, to filtration \cite{Araujo2006}, combustion, deposition of sprays, jet mills, fluidized beds, pneumatic transport of powders, etc.  Together with transport of particles naturally mixing and segregation occurs. In many industrial applications a fine control over these two effects is essential \cite{muzzio2002}. To achieve this, different procedures have been developed. Mechanical mixers like tumbling mills, drum mixers or impeller type mixer are widely used. Another successfully applied technique consists of using jet nozzles that cause bubble formation in fluidized beds. In this situation the mixing of particles is driven by inhomogeneities in the particle density.

In many industrial applications pneumatic conveyor systems are used to transport particles from one process section to the next. Such conveyor systems usually consist of a channel wherein a turbulent air flow is used to fluidize and transport particles. In these systems the particle densities can be considered to be homogeneous enough to avoid mixing driven by density inhomogeneities as mentioned above. Still mixing can be observed and this effect could be used in industrial applications in the future. In this situation the dispersion of particles is determined entirely by the fluid flow. In flows with a high Reynolds number the statistics of velocities and accelerations present inside the turbulent flow are well known from experiments \cite{laporta:1017}. We study the influence of these intrinsic fluctuations on the mixing of systems of constant particle density. For this we couple a particle simulation to a stochastic model for the fluid velocity fluctuations \cite{reynolds:084503} that reproduces well the observed statistics of the fluid velocity and acceleration. We investigate the influence of the strength of the turbulence on the mixing.

Different approaches  for numerical simulations of different kind of mixers have been developed (see e.g. Bertrand \textit{et al.} \cite{bertrand2005} for a review). A popular type of models simulates the particles individually and treats occurring collisions explicitly \cite{rosato1986,brady2006,cundall1979}. For simulating fluidized beds one usually couples a particle based method like the Discrete Element Method (DEM) \cite{cundall1979} with a fluid model. A good review of current techniques has been published by Deen \textit{et al.} \cite{Deen2007}. The basic ideas for simulating bubble formation were first introduced by Tsuji \textit{et al.} \cite{tsuji79} and Hoomans \textit{et al.} \cite{hoomans1996} or was more recently also used in \cite{McNamara2000}.

A large variety of models for fluid-particle coupling have been proposed and a good overview of their current state can be found in the review of Zhu \textit{et al.} \cite{zhu2007}. One important class of techniques consists in using direct numerical simulation (DNS) \cite{choi2001} or Lattice-Boltzmann (LB) \cite{zhang1999} to solve the Navier-Stokes equations and couple the resulting fluid velocity field to the particle motion by an empirical drag law \cite{Li2003}. Unfortunately these techniques are limited to only moderate values of Reynolds numbers. For higher Reynolds numbers one has to relay on turbulence models like Large-Eddy simulations, mixing length models or other turbulence models. These models are usually not able to resolve the smallest fluctuations of the fluid field but rather calculate a (temporal or spatial) average of the fluid velocity field \cite{pope:2000}.

\begin{figure}[tp]
	\includegraphics[width=\columnwidth]{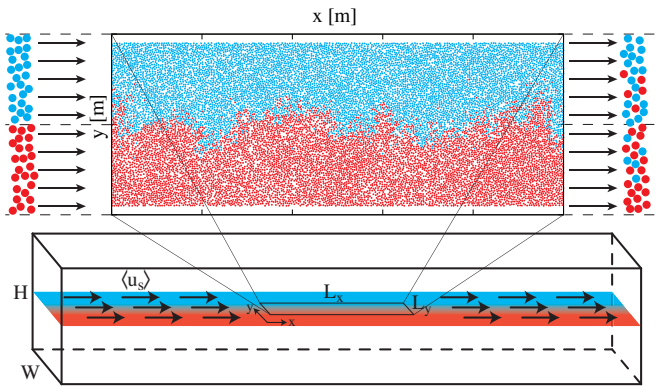}
	\caption{\label{fig:system}(color online) Schematic side view of a rectangular channel of width $W$ and height $H$. Two species of particles are transported through the channel by a turbulent flow of mean velocity $\left< \vect{u}_{\ind{s}}\right>$. The simulation concentrates on a small slice of size $L_{x}\times L_{y}$ located at the center of the channel. The two species enter the simulation on the left side and leave it again on the right side. During this displacement the particles are mixed due to the turbulent flow. The upper part of the figure shows a snap-shot of the system seen from above for $R_{\lambda}=1000$ and $\rho_{v}=0.5$.}
\end{figure}

In this paper we investigate the influence of the intrinsic fluctuations of the turbulent velocity field on the mixing. We study two species of spherical particles flowing through a rectangular pipe as sketched in Fig.~\ref{fig:system}. We neglect all influences from the wall on the fluid velocity field but instead use a stochastic model \cite{reynolds:084503} that reproduces well the measured acceleration and the velocity statistics of tracer particles in fully developed turbulence by La Porta \textit{et al.} \cite{laporta:1017}. We start first with a two dimensional system and extend it to three dimensions. The influence of different Reynolds numbers as well as different densities is analyzed.

\section{\label{sec:model}The Model}
Our model is based on a two-dimensional DEM simulation \cite{ludig:785} in a rectangular domain. The particles are non-deformable disks with radius $r_{i}$ and density $\rho_{i}$. The equations of motion for a particle $i$ are given by
\begin{equation}
	m_{i}\frac{\mathrm{d}^{2}}{\mathrm{d}t ^{2}}\vect{x}_{i}=\vect{F}_{i},
\end{equation}
where $m_{i}$ and $\vect{x}_{i}$ are the particle mass and position and $\vect{F}_{i}$ is the sum of all forces acting on the particle. In our model $\vect{F}_{i}$ is the sum of a collision $\vect{F}_{i}^{\ind{(c)}}$ and a drag force $\vect{F}_{i}^{\ind{(d)}}$ due to the fluid in the channel,

\begin{equation}
	\vect{F}_{i} = \vect{F}_{i}^{\ind{(c)}}+\vect{F}_{i}^{\ind{(d)}}.
\end{equation}

Given $\vect{F}_{i}$, the equations of motion are integrated using the standard Verlet algorithm.

\subsection{Calculation of the collision force}
To model the collision force between two particles with positions $\mathbf{x}_{i}$ and $\mathbf{x}_{j}$, we first calculate their overlap
\begin{equation}
	\delta_{ij}=(r_{i}+r_{j})-\left|\vect{x}_{i}-\vect{x}_{j}\right|,
\end{equation}
If $\delta_{ij}$ is positive, a repulsive force 

\begin{equation}
	\vect{F}_{ij}^{\ind{(c)}}=(k\delta_{ij}+\gamma_{0}v_{ij}^{\ind{(n)}})\, \vect{n}_{ij}
\end{equation}
in the normal direction
\begin{equation}
	\mathbf{n}_{ij}=\frac{\mathbf{x}_{i}-\mathbf{x}_{j}}{\left| \mathbf{x}_{i}-\mathbf{x}_{j} \right|}
\end{equation}
acts on particle $i$. Here $k$ is a spring stiffness, $\gamma_{0}$ a viscous damping coefficient and $v_{ij}^{\ind{(n)}}$ the relative velocity of the particles in the normal direction
\begin{equation}
	v_{ij}^{\ind{(n)}} = -(\mathbf{v}_{i}-\mathbf{v}_{j})\cdot \mathbf{n}_{ij}.
\end{equation}

\subsection{Calculation of the drag forces\label{sec:drag}}
Next we have to specify how the drag force is calculated. Given a (local) fluid velocity $\mathbf{u}_{\ind{s}}$ we follow Bini and Jones \cite{bini:035104} by assuming the simplified expression for the acceleration of particle $i$ due to the flow
\begin{equation}
	\vect{a}_{i}=\frac{\vect{u}_{\ind{s}}-\vect{v}_{i}}{\tau},
\end{equation}
where $\vect{v}_{i}$ is the particle velocity and $\tau$ is a relaxation time. In a laminar flow one can choose $\tau$ to be constant resulting in a drag force $\left|\vect{F}_{i}^{\ind{(d)}}\right|$ acting on particle $i$ proportional to the velocity difference
\begin{equation}
	\left|\vect{F}_{i}^{\ind{(d)}}\right|\propto \left| \mathbf{u}_{\ind{s}} - \mathbf{v}_{i} \right|.
\end{equation}
In the turbulent case one generally assumes a quadratic dependence of the drag force on the velocity difference
\begin{equation}
	\left|\vect{F}_{i}^{\ind{(d)}}\right|\propto \left| \mathbf{u}_{\ind{s}} - \mathbf{v}_{i} \right|^2.
\end{equation}
To achieve this an appropriate expression \cite{bini:035104} for $\tau$ is
\begin{equation}
	\tau^{-1}=\frac{3}{8}\frac{\rho_{f}C_\ind{D}}{\rho_{i}r_{i}}\left| \vect{u}_{\ind{s}} - \vect{v}_{i} \right|,
\end{equation}
where $\rho_{f}$, $\rho_{i}$, $r_{i}$ and $C_{\ind{D}}$ are the fluid density, the particle density, the particle radius and the drag coefficient, respectively. For spherical particles at high Reynolds numbers $C_\ind{D}\approx 0.45$.

At this point it is important to note that the fluid is not influenced by the particles. In a more detailed simulation one would need to include a feedback on the fluid due to the energy transfer from the fluid to the particles.

\subsection{Determination of the turbulent flow\label{sec:turbulence}}
To complete our model we need to define the turbulent flow velocity $\vect{u}_{\ind{s}}$. We first split $\vect{u}_{\ind{s}}$ into the mean stream velocity $\left< \vect{u}_{\ind{s}} \right>$ and a stochastic part $\vect{u}_{\ind{t}}$:
\begin{equation}\label{eq:flow_decomposition}
	\vect{u}_{\ind{s}}=\left< \vect{u}_{\ind{s}} \right>+\vect{u}_{\ind{t}}.
\end{equation}
Calculating the mean flow velocity $\left< \mathbf{u}_{s} \right>$ is rather easy and would also account for the wall effects. For example one could apply a turbulent viscosity model \cite{pope:2000}, like the mixing length model or the widely used $k-\varepsilon$ model, to determine $\left< \mathbf{u}_{s} \right>$.

Determining the stochastic part $\mathbf{u}_{t}$ of the flow is more difficult. Several measurements \cite{laporta:1017,mordant:214501,voth:121} have shown a highly non-Gaussian behavior of the Lagrangian acceleration distribution of fluid particles. Sawford \cite{sawford:1577} introduced a system of stochastic differential equations (SDE) for the time evolution of the velocity and acceleration of a fluid particle. This model correctly reproduces the observed Gaussian distribution of the velocities. Unfortunately it fails to show the highly non-Gaussian distributions for the acceleration. To overcome this problem Pope and Chen \cite{pope:1437} introduced an Uhlenbeck-Ornstein process for the evolution of the (now fluctuating) energy dissipation rate $\varepsilon$. Reynolds \cite{reynolds:084503} combined these ideas and formulated a system of SDEs for the logarithm of the dissipation rate $\chi=\log(\varepsilon/\left< \varepsilon \right>)$, and the components of the accelerations $a_{t}$ and velocities $u_{t}$ of tracer particles in fully developed turbulence. This model reproduces well \cite{Mordant:245} the observed probability density functions as well as the autocorrelation functions of the velocity and acceleration:
\begin{subequations}\label{eq:sde}
	\begin{eqnarray}
		d\chi & = & -\left( \chi - \left< \chi \right> \right)T_{\chi}^{-1}dt+\sqrt{2\sigma_{\chi}^{2}T_{\chi}^{-1}}d\xi_1\label{eq:sde_chi}\\
		da_{t} & = & -\left( T_{L}^{-1} + t_{\eta}^{-1} -\sigma_{a_{t}|\varepsilon}^{-1} \frac{d\sigma_{a_{t}|\varepsilon}}{dt} \right) a_{t} dt\nonumber\\*
		& & -T_{L}^{-1} t_{\eta}^{-1} u_{t} dt\nonumber\\*
		& & + \sqrt{2\sigma_{u}^2\left( T_{L}^{-1} + t_{\eta}^{-1} \right)T_{L}^{-1} t_{\eta}^{-1} }d\xi_2\label{eq:sde_a}\\
		du_{t} & = & a_{t}dt\label{eq:sde_u}.
	\end{eqnarray}
\end{subequations}
In Eq.~(\ref{eq:sde_chi}) the variance of $\chi$ is approximated by $\sigma_{\chi}^2=-0.354+0.289\log R_{\lambda}$ \cite{yeung531}, its mean value is given by $\left< \chi \right> = -0.5 \sigma_{\chi}^2$, the relaxation time scale $T_{\chi}=2\sigma_{u}^2/(C_0\left< \varepsilon \right>)$ and $\left< \varepsilon \right>$ is the mean energy dissipation rate. The Reynolds number at the Taylor microscale $R_\lambda$ is related to the Reynolds number by $R_{\lambda}=\sqrt{15\mathrm{Re}}$.  In Eq.~(\ref{eq:sde_a}) we have the energy-containing scale $T_{L}=2\sigma_{u}^2/(C_{0}\varepsilon)$, the energy-dissipation scale $t_{\eta}=C_{0}\nu^{1/2}/(2a_{0}\varepsilon^{1/2})$, the conditional acceleration variance $\sigma_{a_{t}|\varepsilon}^{2}=a_0\varepsilon^{3/2}\nu^{1/2}$, two universal Lagrangian velocity structure constants $a_{0}=3.3$ and $C_{0}=7.0$, the kinematic viscosity $\nu$ and the velocity variance $\sigma_{u}^{2}$. The values of $a_{0}$ and $C_{0}$ are determined by demanding consistency with Kolmogorov's (1941) hypothesis \cite{monin1975} and fitting them to experimental data \cite{reynolds:084503}. Finally there are two independent Wiener processes, i.e. Gaussian distributed random numbers, $d\xi_1$ and $d\xi_2$ with zero mean and variance $dt$.

With this model we are able to calculate each component of the turbulent velocity $\vect{u}_{\ind{t}}$. We combine one equation for $\chi$ and two equations for $a_{t}$ and $u_{t}$ to get a two-dimensional vector $\vect{u}_{\ind{t}}$ for the velocity of a tracer particle. We then generate a set of these tracer particles and evolve them in time according to Eq.~\ref{eq:sde}. When a new particle enters the system, we randomly chose a tracer particle out of this set and use this (tracer) particle to calculate the stochastic forces acting on the (real) particle until the particle leaves the system. This means that we add the mean velocity $\left< \vect{u}_{\ind{s}} \right>$ and the stochastic part $\vect{u}_{\ind{t}}$ from the assigned tracer particle to get a ``stream line'' which we then use to calculate the drag force $\vect{F}_{i} ^{\ind{(d)}}$.

\section{Simulation Set-up\label{sec:setup}}
We applied the model described in the last section to a system with two types of particles flowing through a channel. A sketch of the system is shown in Fig.~\ref{fig:system}. Both types of particles have the same radius $r=10^{-5}\text{ m}$ and density $\rho=3.0\cdot 10^{3}\text{ kg m}^{-3}$ (but a different ``color'' to distinguish them). The spring stiffness and viscous damping coefficient are set to $k=1.0\ \mathrm{kg}\,\mathrm{s}^{-2}$ and $\gamma_{0}=1.0\cdot 10^{-6}\ \mathrm{kg}\,\mathrm{s}^{-1}$ and kept constant for all simulations. The fluid inside the channel has a kinematic viscosity $\nu =1.64\cdot 10^{-5}\text{ m}^{2}\text{s}^{-1}$. The channel has a width $W$ and height $H$. In our case we have $W=0.2\text{ m}$ and $H=0.358\text{ m}$. To get a two dimensional system we first cut out a horizontal slice from the channel. We further want to simplify the setup and study the basic effects of the stochastic part in Eq.~(\ref{eq:flow_decomposition}). Therefore we concentrate on a small region in the center of the channel with dimensions $L_{x}=0.025\text{ m}$ and $L_{y}=0.002\text{ m}$. Next we set the mean velocity to be constant throughout the whole system and pointing into the positive x-direction, $\left< \vect{u}_{\ind{s}} \right>=\left( \left<u_{\ind{s}}\right>,0 \right)$. This of course ignores wall effects, because the mean flow profile in a channel is not constant and drops to zero at the boundaries. But as mentioned before our simulation focuses only on the region in the center of a large channel where the mean velocity is approximately constant. These assumptions influence the choice of our boundary conditions which will be discussed later.

The magnitude of $\left< u_{\ind{s}} \right>$, as well as the mean energy dissipation rate $\left< \varepsilon \right>$ and the velocity fluctuations $\sigma_{u}^2$ depend on the Reynolds number. The values used in our calculations are listed in Tab.~\ref{tab:parameters}. These parameters were determined in the following way:
\begin{enumerate}
	\item For pipe flow the Reynolds number at the Taylor microscale \cite{pope:2000} can be calculated by
	\begin{equation}\label{eq:Rpipe}
		R_{\lambda}=\sqrt{15\frac{\left<u_{\ind{s}}\right>}{\nu}D_{\ind{H}}}.
	\end{equation}
	Here $D_{\ind{H}}$ is the hydraulic diameter. For a rectangular duct of width $W$ and height $H$ this is given by
	\begin{equation}
		D_{H}=\frac{2WH}{W+H}.
	\end{equation}
	With this expression one can calculate $\left<u_{\ind{s}}\right>$ for a given Reynolds number $R_{\lambda}$.
	\item In fully developed turbulence there is an alternative expression for the Reynolds number \cite{pope:2000}:
	\begin{equation}\label{eq:Rturbulence}
		R_{\lambda}=\frac{\sigma_{u}\lambda}{\nu}=\sqrt{\frac{15}{\left<\varepsilon\right>\nu}}\sigma_{u}^{2},
	\end{equation}
	where $\lambda=\sqrt{15\nu\sigma_{u}^{2}/\left<\varepsilon\right>}$ is the Taylor microscale. For a fixed Reynolds number $R_{\lambda}$ this gives us a relation between the mean energy dissipation rate $\left<\varepsilon\right>$ and the velocity fluctuations $\sigma_{u}$. Another relation between these two quantities is given by
	\begin{equation}
		\left<\varepsilon\right>=\frac{\sigma_{u}^{3}}{L},
	\end{equation}
	where $L$ is a ``typical'' length scale. We assume that $L$ is of the same order as $D_{\ind{H}}$, i.e. $L\sim D_{\ind{H}}$. This allows us to calculate values for $\left<\varepsilon\right>$ and $\sigma_{u}$.
	\item The Stokes number, quantifying the effect of inertia on the suspended particles, can be calculated \cite{Collins2004} as
	\begin{equation}
		\mathrm{St}=\frac{\tau_{i}}{\tau_{\eta}}
	\end{equation}
	where $\tau_{i}$ is the particle relaxation time
	\begin{equation}
		\tau_{i}=\frac{4 r_{i}^{2} \rho_{i}}{18\nu \rho_{f}}
	\end{equation}
	and $\tau_{\eta}$ is the Kolmogorov time scale
	\begin{equation}
		\tau_{\eta}=\sqrt{\frac{\nu}{\left< \epsilon \right>}}.
	\end{equation}
\end{enumerate}

\begin{table}[t]
	\caption{\label{tab:parameters}List of Reynolds number dependent parameters in our simulations.}
	\begin{ruledtabular}
		\begin{tabular}{ccccc}
			$R_{\lambda}$ & $\mathrm{St}$ & $\left< u_{s} \right>$ & $\left< \varepsilon \right>$ & $\sigma_{u}^2$\\
			$-$ & $-$ & $\mathrm{m}\mathrm{s}^{-1}$ & $\mathrm{m}^{2}\mathrm{s}^{-3}$ & $\mathrm{m}^{2}\mathrm{s}^{-2}$\\
			\hline
			$50$ & $0.0019$ & $0.01065$ & $4.708\cdot 10^{-6}$ & $0.0001134$\\
			$100$ & $0.0153$ & $0.04260$ & $3.013\cdot 10^{-4}$ & $0.001815$\\
			$150$ & $0.0516$ & $0.09586$ & $3.432\cdot 10^{-3}$ & $0.009189$\\
			$200$ & $0.1223$ & $0.1704$ & $1.928\cdot 10^{-2}$ & $0.02904$\\
			$300$ & $0.4127$ & $0.3834$ & $2.197\cdot 10^{-1}$ & $0.1470$\\
			$400$ & $0.9781$ & $6.817$ & $1.234\cdot 10^{0}$ & $0.4647$\\
			$500$ & $1.9105$ & $1.065$ & $4.708\cdot 10^{0}$ & $1.134$\\
			$600$ & $3.3016$ & $1.534$ & $1.406\cdot 10^{1}$ & $2.352$\\
			$700$ & $5.2426$ & $2.088$ & $3.545\cdot 10^{1}$ & $4.358$\\
			$800$ & $7.8257$ & $2.727$ & $7.899\cdot 10^{1}$ & $7.434$\\
			$900$ & $11.1412$ & $3.451$ & $1.601\cdot 10^{2}$ & $1.191$\\
			$1000$ & $15.2840$ & $4.260$ & $3.013\cdot 10^{2}$ & $18.15$
		\end{tabular}
	\end{ruledtabular}
\end{table}

Next we have to specify the boundary conditions of our system. New particles enter the system at the left side and leave it again at the right side (See Fig.~\ref{fig:system}). The boundary conditions in the transverse direction are more complex: Particles are elastically bounced back by reflecting their velocity vector such that the vertical component points back into the system. This leaves the total kinetic and potential energy of the particles constant. Additionally, we exchange the to the particle assigned tracer particle that is used to calculate the stochastic part of the fluid velocity. This is done by drawing uniformly distributed a tracer particle out of our set of all tracer particles. Such a boundary condition is reasonable because we concentrate only on a small region in the middle of the channel and our boundaries are not the walls of this channel. One can imagine that whenever a particle hits our boundaries, in reality it would leave our domain of simulation. So, to keep the (local) particle number constant, we think of the reflected particle as a new particle that would enter our simulation domain at the same place and time where the old particle left.

Finally we have to specify the density of our system by fixing the volume fraction $\rho_{v}$ that is occupied by particles
\begin{equation}
	\rho_{v}=\frac{(n_{1} \cdot V_{1}+ n_{2} \cdot V_{2})}{L_{x}\cdot L_{y}}.
\end{equation}
Here $n_{k}$ and $V_{k}$ denote the number of particles of type $k$ and their respective volume. This determines how many particles are on average in the system. Using the mean stream velocity we can estimate how many particles leave the system and therefore how many new particles we have to insert at the opposite side into the simulation to keep $\rho_{v}$ constant. We divide the left boundary of the system in two separate equally sized intervals, insert the one species into the right and the other species into the left interval. New particles are placed in such a way that they do not overlap with existing ones. A section of the system at the end of the channel is shown in the upper part of Fig.~\ref{fig:system}.

\section{\label{sec:results}Results}
To quantify the quality of mixing, we cut out vertical slices of a certain width from the system. These slices we divide into horizontal intervals and count how many particles of which type are in each interval. Averaging over many time-steps we get a time-averaged relative density for each interval. To be more specific, let's look at a slice located at $x_{s}$. We divide this slice into $n_{s}$ horizontal intervals. This gives us $n_{s}$ small boxes $\mathcal{B}_{i}$, centered at $\left( x_{s}, \frac{L_{y}}{n_{s}}\cdot (i+0.5) \right)$. During the simulation we count the number of particles of each type, namely $N_{i,1}$ particles of type 1 and $N_{i,2}$ particles of type 2 in the box $\mathcal{B}_{i}$. From this we can calculate the relative densities
\begin{equation}
	\rho_{i,k}=\frac{N_{i,k}}{N_{i,1}+N_{i,2}},\qquad k=1,2.
\end{equation}

\begin{figure}[bp]
	\includegraphics[width=\columnwidth]{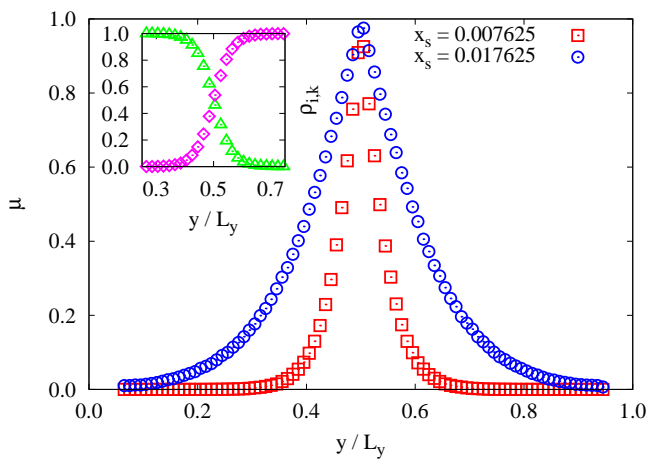}
	\caption{\label{fig:mix}(color online) Example of the behavior of $\mu_{i}$ along two slices with $R_{\lambda}=1000$ and $\rho_{v}=0.2$. The inset shows the behavior of the relative densities of the two species in the center of the channel at $x_{s}=0.007625$ with $R_{\lambda}=1000$ and $\rho_{v}=0.2$.}
\end{figure}

The behavior of these relative densities for the two species of particles in the center of the channel along one slice is shown in the inset of Fig.~\ref{fig:mix}.

\subsection{Growth of the mixing zone}\label{sec:mixingZone}
Now we will quantify the size of the mixing zone. Therefore we calculate
\begin{equation}
	\mu_{i}=1-\left| \rho_{i,1}-\rho_{i,2} \right|
	\label{eq:relDens}
\end{equation}
to get a quantity which is close to zero in regions where the mixing is bad, increases the more the particles are mixed and approaches unity when the two species are well mixed. Examples for the behavior of $\mu_{i}$ along two slices are shown in Fig.~\ref{fig:mix}. 

\begin{figure}[tp]
	\includegraphics[width=\columnwidth]{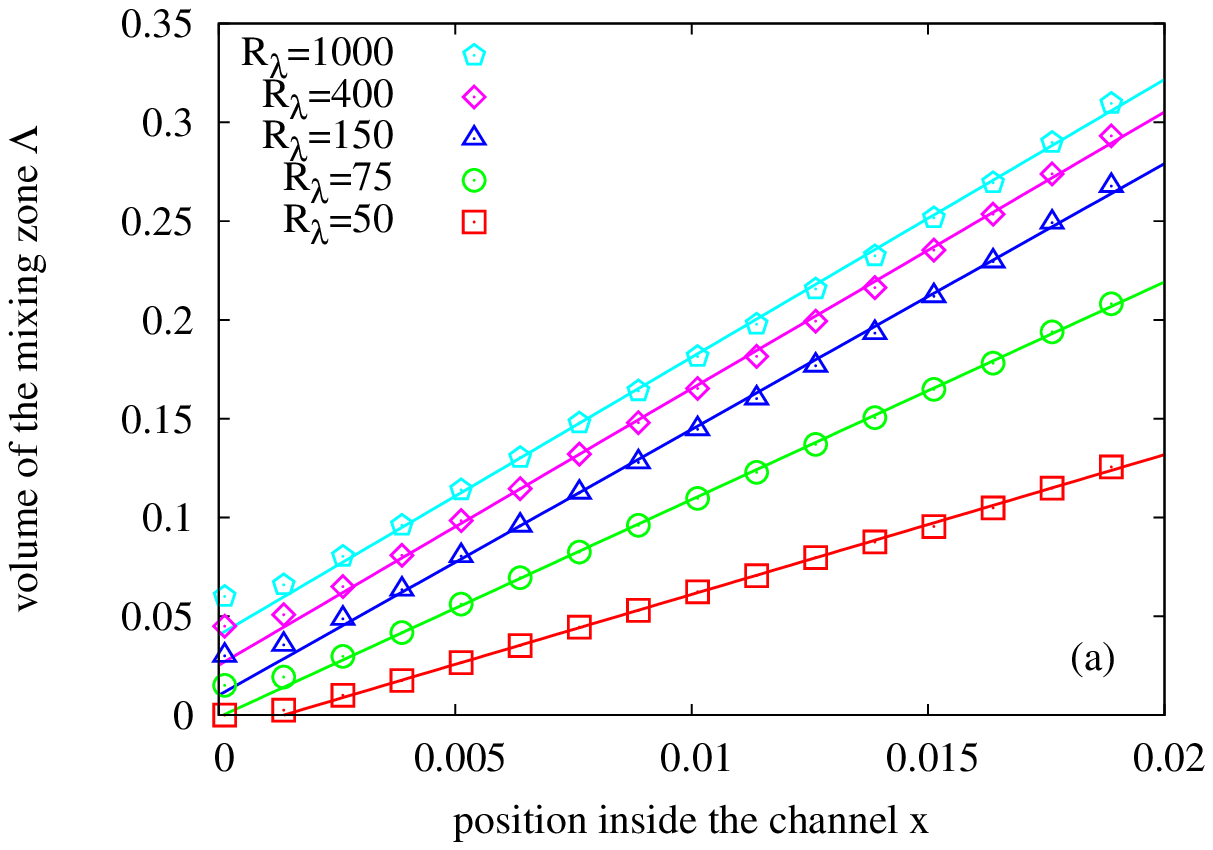}
	\includegraphics[width=\columnwidth]{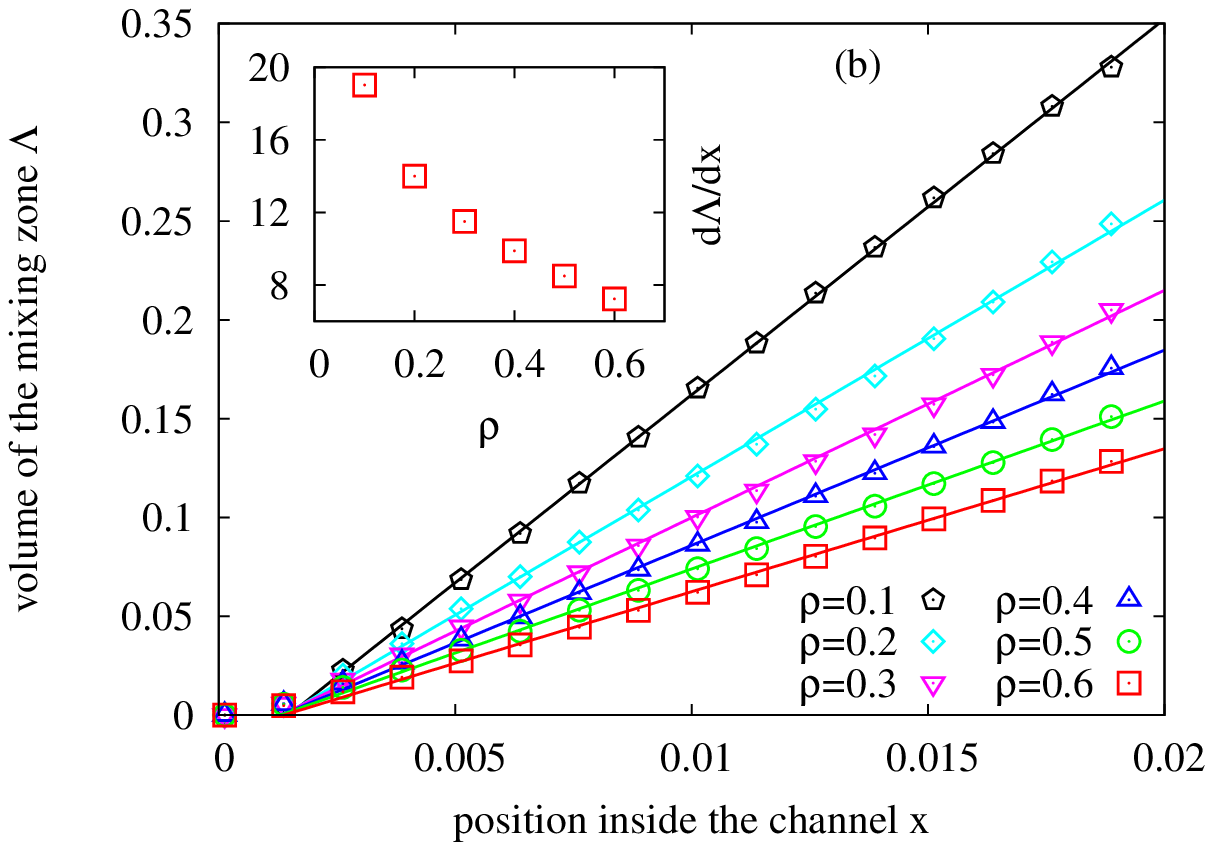}
	\caption{\label{fig:lambda}(color online) Behavior of the size $\Lambda$ of the mixing zone in two dimensions for different Reynolds numbers and densities: (a) Behavior of $\Lambda(x)$ for constant density $\rho_{v}=0.2$ and different Reynolds numbers $R_{\lambda}$. One observes a linear increase of the mixing zone along the channel. The mixing gets faster with increasing Reynolds numbers until it reaches a limit, i.e. the slopes $d\Lambda / dx$ of the linear fits stay constant. Curves are shifted vertically to better distinguish the individual curves. (b) Behavior of $\Lambda(x)$ for constant Reynolds number $R_{\lambda}=800$ and different densities. Systems with lower densities exhibit faster mixing. The inset shows the growth rate $d\Lambda/d\rho$ of the mixing zone.}
\end{figure}

\begin{figure}[tp]
	\includegraphics[width=\columnwidth]{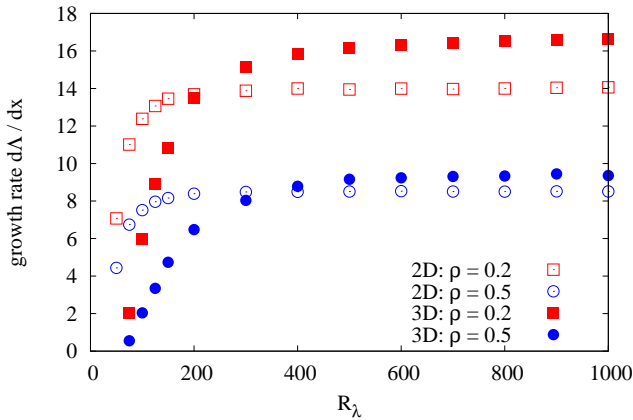}
	\caption{\label{fig:increase}(color online) Dependence of the slope $d\Lambda /dx$ on the Reynolds number $R_{\lambda}$. Above a certain Reynolds number $R_{c}\approx 300$ the slope stays almost constant. Empty symbols represent the results in 2D and filled symbols are the ones for 3D, discussed in section \ref{sec:3D}.}
\end{figure}

We now calculate the area $\Lambda$ below the measured points $\mu_{i}$ to get a quantity that specifies the size of the mixing zone
\begin{equation}
	\Lambda = \sum_{i=1}^{n_{s}}\mu_{i}.
\end{equation}
We can calculate $\Lambda$ for every slice and find a linear increase with increasing length of the channel (see Fig.~\ref{fig:lambda}~(a)). By altering the Reynolds number we find an increase of the slope of $\Lambda(x_{s})$ with increasing Reynolds number. Above a certain Reynolds number $R_{c}$ the quality of mixing does not increase any more, i.e. for $R_{\lambda}\rightarrow \infty$ the slope $d\Lambda / dx$ approaches a constant value. This result is shown in Fig.~\ref{fig:increase}. In our simulation $R_{c}\approx 300$. This result has implications on industrial applications where mixing in turbulent flows is of key importance.

If we fix the Reynolds number and alter the density, we find a faster mixing for lower densities  (see Fig.~\ref{fig:lambda}~(b)). The dependence of the growth rate of the mixing zone on the density is shown in the inset of Fig.~\ref{fig:lambda}~(b). With our current implementation it is possible to reach densities of about $0.6$, because particles are not allowed to overlap when they enter the system. By using another way of inserting particles into the system it may be possible to reach even higher densities close to the value for random close packing $\rho_{vc}\approx0.84$. Because particles can be elastically deformed, we expect that even at densities close to random close packing we can still observe mixing and the growth rate of the mixing zone goes to a finite value. This tendency can be seen in the inset of Fig.~\ref{fig:lambda}~(b).

\subsection{Lateral dispersion}
\begin{figure}[bp]
	\begin{center}
		\includegraphics[width=\columnwidth]{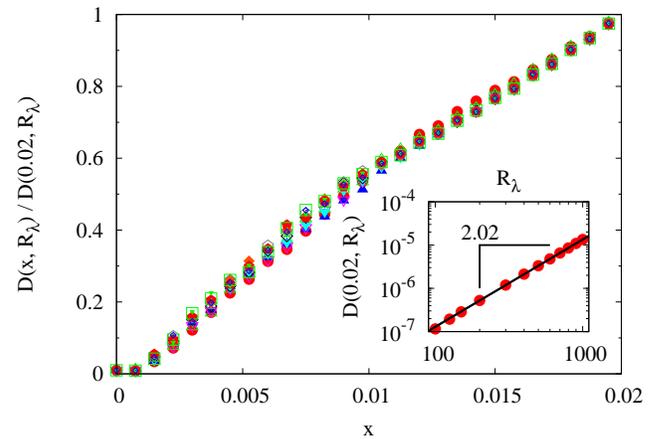}
	\end{center}
	\caption{(color online) Data collapse of the dispersion coefficients $D(x,R_{\lambda})$ for $\rho_{v}=0.2$. For every Reynolds number $R_{\lambda}$ the corresponding curve along the channel was rescaled with the dispersion coefficient at $x=0.02$. The inset shows the quadratic dependence of the values $D(x=0.02, R_{\lambda})$ on the Reynolds number $R_{\lambda}$.}
	\label{fig:dispersion1}
\end{figure}

As an alternative to measuring the size of the mixing zone we use a dispersion relation to quantify the rate of mixing. The motion of tracer particles can be characterized \cite{Naumann1983} by a convection-diffusion equation

\begin{equation}
	\frac{\partial \rho_{k}(\vect{x},t)}{\partial t} + \vect{u}_{\ind{s}}(\vect{x})\cdot\nabla\rho_{k}(\vect{x},t)=\nabla\cdot (D(\rho_{k},\vect{x})\nabla\rho_{k}(\vect{x},t)),
	\label{eq:cde}
\end{equation}

where $\rho_{k}(\vect{x},t)$ is the relative density of particles of species $k$ at position $\vect{x}$ and time $t$, $\vect{u}_{\ind{s}}(\vect{x})$ the fluid velocity and $D(\rho_{k},\vect{x})$ the coefficient of dispersion. Eq.~\ref{eq:cde} can be simplified as follows:
The first term vanishes by assuming a steady state 
\begin{equation}
	\frac{\partial \rho_{k}(\vect{x},t)}{\partial t}=0.
\end{equation}
The second term simplifies by replacing the fluid velocity by the mean fluid velocity
\begin{equation}
	\vect{u}_{\ind{s}}(\vect{x})=\left< \vect{u}_{\ind{s}} \right> = \left( \left< u_{\ind{s}} \right>,0 \right)
\end{equation}
The right hand side simplifies by assuming
\begin{equation}
	D(\rho_{k},\vect{x})=D=\mathrm{const.}
\end{equation}
and by using according to Ref.~\cite{Harleman1963}
\begin{equation}
	\frac{\partial^{2}\rho_{k}(\vect{x},t)}{\partial x^{2}}\ll \frac{\partial^{2}\rho_{k}(\vect{x},t)}{\partial y^{2}}.
\end{equation}
Therefore Eq.~\ref{eq:cde} simplifies to

\begin{equation}
	\left<u_{\ind{s}}\right>\frac{\partial \rho_{k}}{\partial x}=D \frac{\partial^{2}\rho_{k}}{\partial y^{2}}.
	\label{eq:lateralCDE}
\end{equation}
If one assumes Heaviside functions as the initial density profiles at $x=0$, the solution of Eq.~\ref{eq:lateralCDE} is given \cite{Harleman1963} by
\begin{equation}
	\rho_{k}=\frac{1}{2}\left( 1 \pm \erf\left( \frac{y}{2 \sqrt{D x/\left< u_{\ind{s}} \right>}} \right) \right)
\end{equation}
where $\erf(x)$ is the Error function. Rewriting this equation leads to
\begin{equation}\label{eq:dispersionSol}
	\erf^{-1}\left( 1-2\cdot \rho_{k} \right) = \pm \frac{1}{2\sqrt{D}} \frac{y}{\sqrt{x/\left< u_{\ind{s}} \right>}}.
\end{equation}
For every position of a slice $x=x_{\ind{s}}$ we can now calculate the left and right hand side of Eq.~\ref{eq:dispersionSol} along the $y$-direction and fit for the dispersion coefficient $D$. In theory this should lead to the same dispersion coefficient for all slices. As an example we show in Fig.~\ref{fig:dispersion1} the results for a system with $\rho_{v}=0.2$. As can be seen $D$ is not constant along the channel and depends on the position $x$ inside the channel and the Reynolds number $R_{\lambda}$. Rescaling the values of $D(x,R_{\lambda})$ by $1/D(x=0.02,R_{\lambda})$ leads to a data collapse of the curves. The inset of Fig.~\ref{fig:dispersion1} shows that the values of $D(0.02,R_{\lambda})$ seem to depend quadratically on the Reynolds number $R_{\lambda}$. The same behavior can be observed for other densities as well. In Fig.~\ref{fig:dispersion1} one may also suspect a crossover between two linear regimes. However, this does not seem to be observable for other densities.

For small values of $x$ the variation of $D$ may be due to the initial conditions of particles entering the system, where all particles have the same velocity $\left< \vect{u_{\ind{s}}} \right>$ pointing in the positive $x$-direction. But because the values of $D$ do not converge towards a constant value it seems likely that the lateral dispersion of particles in our system cannot be described by normal dispersion. To conclusively judge whether our system shows anomalous dispersion or not further investigations e.g. of the behavior of the mean squared displacement of one particle over time would be needed.

Even the dispersion coefficients being not constant throughout the system, we tried to compare them with the turbulent diffusivity of Taylor \cite{Taylor1921,Naumann1983} defined by
\begin{equation}
	D_{\ind{E}}=\sigma_{u}^{2}T_{L}.
	\label{eq:DE}
\end{equation}
Inserting the formula for the Lagrangian integral time scale $T_{L}$ given in Sec.~\ref{sec:turbulence} leads to
\begin{equation}
	D_{E}=\frac{2\nu}{15 C_{0}}R_{\lambda}^{2}.
	\label{eq:DE2}
\end{equation}
This explains the observed quadratic dependence of $D$ on $R_{\lambda}$. Inserting the values for the viscosity $\nu$ and model parameter $C_{0}$ results in values for $D_{E}$ of the order $10^{-4}$ for small Reynolds numbers to $10^{-1}$ for large ones. These values are clearly orders of magnitude above the values observed in our simulations.

\subsection{Particle-particle and fluid-particle interactions}
The motion of particles in our system is determined by two types of interactions: The particle-particle collisions and the drag force from the fluid. The initial conditions of particles entering our system are chosen such that these particles do not overlap and their velocities are equal to the mean flow velocity $\left< \vect{u}_{\ind{s}} \right>$. If we remove the stochastic part of $\vect{u}_{\ind{s}}$ particles would just fly along straight lines and one would not observe any mixing at all. By adding random components to the initial velocity of the particles the mixing would be similar to that in a granular gas. In our simulation the particles all have the same velocity when entering the system and variations of the initial velocity are introduced by the stochastic forces described in Sec.~\ref{sec:drag} and \ref{sec:turbulence}. To estimate the influence of the particle-particle and fluid-particle interactions we can compare the mean free time between collisions in a granular gas \cite{Chapman:1990}
\begin{equation}
	\tau_{g}=\frac{\pi r_{i}}{2\sqrt{2}\rho_{v}\left< u_{\ind{s}}\right>}
\end{equation}
with the Kolmogorov time $\tau_{\eta}$ of the turbulent part $\vect{u}_{\ind{t}}$ of the fluid velocity. The mean free time $\tau_{g}$ is related to the mean free path $\lambda$ of a granular gas by dividing $\lambda$ by the mean particle velocity. Inserting the parameters for our simulations shows that the Kolmogorov time is about $10-1000$ times larger than $\tau_{g}$. This means that many particle collisions occur between changes of the fluid velocity. In this case the Stokes number plays an important role. Tab.~\ref{tab:parameters} shows that for $R_{\lambda}\lesssim 400$ the Stokes number is smaller than one which means that particles follow the fluctuations in the fluid velocity $\vect{u}_{\ind{s}}$ rather closely and the particle collisions can be seen as small perturbations only. Therefore the actual values of the spring stiffness $k$ and viscous damping coefficient $\gamma_{0}$ should not have a significant influence on our results in this range of Reynolds numbers, even for the limiting case of hard spheres ($k\rightarrow \infty$). However, $\mathrm{St}\gtrsim 1$ for $R_{\lambda}\gtrsim 400$ which means that particles will be increasingly less influenced by changes in the fluid velocity and particle-particle collisions become more important. This may be a reason for the saturation of the mixing rate for large Reynolds numbers observed in Sec.~\ref{sec:mixingZone}. Considering variations of $k$ and $\gamma_{0}$, in particular the case of hard spheres, may have an influence on this asymptotic value.

It is also worth noting that numerous publications \cite{Squires1991,Bec2005, Bec2007, IJzermans2009} report that particles in turbulent flows tend to concentrate in regions of low vorticity and high strain rate. Bec \textit{et al.} \cite{Bec2005} showed that this effect of preferential concentration can be observed for heavy particles and Stokes numbers $\mathrm{St}\lesssim 1$. From Tab.~\ref{tab:parameters} one can see that this criterion is fulfilled for $R_{\lambda}\lesssim 400$ and therefore clustering of particles should be observable in our simulations. Indeed when looking at a system with zero mean flow velocity $\left< u_{\ind{s}} \right>=0$ and periodic boundary conditions clustering of particles was observed. In the systems studied in this paper one may recognize such clustering at the end of the channel. The reason for this less obvious occurrence of preferential concentration is probably that the initially homogeneous distributed particles leave the system again before a clear clustering can be observed. 

\section{\label{sec:3D}Extension to three dimensions}
We also extended our model to three dimensions. Most parts of the two-dimensional model carry over to the three-dimensional one in a straight forward way. Now, particles become spheres instead of disks. The collision and the drag force are calculated exactly the same way by just replacing two-dimensional vectors by their corresponding three-dimensional ones. The stochastic part $\vect{u}_{\ind{t}}$ of the flow velocity $\vect{u}_{\ind{s}}$ is calculated by adding a third component of $a_{t}$ and $u_{t}$ to the computation. Finally we extend our formerly two-dimensional domain of simulation in the $z$-direction by a length $L_{z}$ to get a three-dimensional rectangular channel located in the center of the big channel. Particles still enter the system at the left side and leave it at the right side. The boundary conditions on the other four sides of the channel are the same as in the two-dimensional model: Particles are bounced back from the wall and their attached tracer particles are exchanged. We studied a system of size $L_{x}=0.01\ \mathrm{m}$, $L_{y}=0.001\ \mathrm{m}$, $L_{z}=0.0003\ \mathrm{m}$ and two different densities $\rho_{v}=0.2$ and $\rho_{v}=0.5$. This system is smaller than the two-dimensional one because the computational cost for a three dimensional computation is much higher. All other parameters stay the same as in two dimensions.

\begin{figure}[tp]
	\includegraphics[width=\columnwidth]{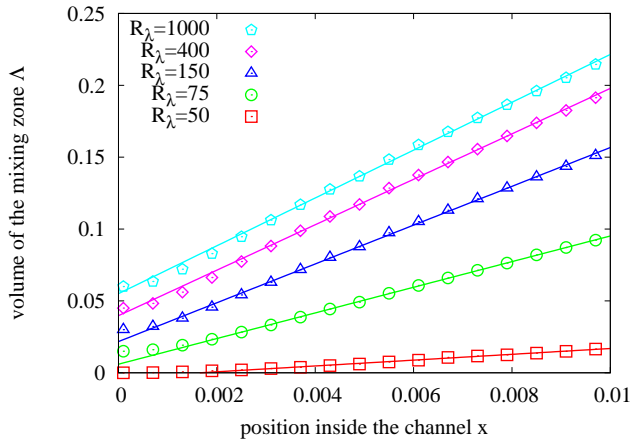}
	\caption{\label{fig:lambda3D}(color online) Behavior of the size $\Lambda$ of the mixing zone in three dimensions for different Reynolds numbers at $\rho_{v}=0.2$. One again observes a linear increase of the mixing zone along the channel. The mixing gets stronger with increasing Reynolds numbers until it reaches a limit, i.e. the slopes $d\Lambda / dx$ of the linear fits stay constant. Curves are shifted vertically to better distinguish the individual curves.}
\end{figure}

To quantify the quality of mixing we used a similar approach as in the two-dimensional case. Again we cut out slices perpendicular to the $x$-axis at different positions $x_{s}$. Every slice we divided into $n_{s,y}$ intervals along the $y$-axis and $n_{s,z}$ intervals along the $z$-axis. This gives us $n_{s,y}\cdot n_{s,z}$ boxes $\mathcal{B}_{i,j}$ centered at $\left( x_{s},\frac{L_{y}}{n_{s,y}}\cdot(i+0.5), \frac{L_{y}}{n_{s,z}}\cdot(j+0.5) \right)$. During the simulation in every box $\mathcal{B}_{i,j}$ we again counted the number of particles of every species, i.e. $N_{i,j,1}$ and $N_{i,j,2}$. Because the system has a translational symmetry along the $z$-axis we calculated
\begin{equation}
	\left< N_{i,k} \right> = \frac{1}{n_{s,z}}\sum\limits_{j}N_{i,j,k},\quad k=1,2.
\end{equation}
This gives us the mean particle number of every species along the $y$-axis. From this we again can calculate the relative densities
\begin{equation}
	\rho_{i,k}=\frac{\left< N_{i,k} \right>}{\left< N_{i,1} \right>+\left< N_{i,2} \right>},\quad k=1,2.
\end{equation}
The qualitative behavior of these densities is the same as in the two-dimensional model shown in the inset of Fig.~\ref{fig:mix}. From here on the calculation is exactly the same as before. We first calculate $\mu_i$, then compute the area $\Lambda$ below these curves and plot $\Lambda$ versus the position $x$ inside the channel. This plot is shown in Fig.~\ref{fig:lambda3D} for $\rho=0.2$ at different Reynolds numbers. Again we find a linear increase with respect to the length of the channel. By changing the Reynolds number we see a similar increase of the slope of $\Lambda(x)$ as in two dimensions. To compare the two and three dimensional models we plot in Fig.~\ref{fig:increase} the slopes of $\Lambda(x)$ for different Reynolds numbers $R_{\lambda}$ and different densities $\rho_{v}$ for the two- and three-dimensional cases. As expected the rate at which the mixing zone grows increases with higher Reynolds numbers and goes to a constant value when $R_{\lambda}\rightarrow\infty$. The mixing is also slower for higher densities. More interesting is the comparison between the mixing rate in two and three dimensions for a fixed density: One would expect, that the mixing is faster in three dimensions because particles have an additional degree of freedom in three dimensions and can move over each other more easily. This behavior we can observe for high Reynolds numbers. But surprisingly for small Reynolds numbers the mixing in three dimensions is actually slower than in two dimensions. Another interesting result is that the nearly constant value of the mixing rate is approached faster in two dimensions than in three.

\section{\label{sec:conclusion}Conclusion}
By combining a molecular dynamics simulation with a PDF method for modeling turbulent flows, we were able to develop a simple model to investigate the mixing of two species of particles inside a channel. The mixing is caused solely by the stochastic force measured in fully developed turbulence. We found that the mixing zone increases linearly with the length of the channel. The mixing is stronger at higher Reynolds numbers and at lower densities. For small Reynolds number the mixing is stronger in two dimensions than in three. We further investigated the lateral dispersion of particles and found hints toward anomalous dispersion.

Further improvements of our model would include a more detailed modeling of the mean fluid flow in the channel to include the wall effects. At the same time the effect of gravity should be included in the simulations. One also should introduce a feedback from the particles back to the fluid and also different sizes and shapes of particles could be studied. Here we investigated non-deformable particles. In future work it would be interesting to study the influence of particle deformations on the mixing. Whether or not anomalous lateral dispersion can be observed remains an open question. Finally investigating situations where the mean free time between particle collisions becomes of the same size as the turbulent velocity autocorrelation time could also significantly change the mixing behavior.

\begin{acknowledgments}
The authors would like to thank the Swiss National Fond Grant NF20021-116050/1. We also thank Peter Kruspan and Andreas G\"o\ss nitzer for interesting discussions.
\end{acknowledgments}

\bibliography{references}

\end{document}